\documentclass[11pt]{article} 
\pdfoutput=1

\usepackage[utf8]{inputenc}
\usepackage[T1]{fontenc} 
\usepackage{lmodern}
\usepackage[english]{babel}           
\usepackage{cancel}
\usepackage[pdftex]{graphicx}
\usepackage{epsfig}
\usepackage{graphicx}
\usepackage{comment}
\usepackage{latexsym}
\usepackage{hyperref}
\usepackage{amsmath}
\usepackage{mathrsfs}
\usepackage[usenames, dvipsnames]{color}
\usepackage{amsbsy}
\usepackage{amssymb}
\usepackage{amsthm}
\usepackage{amsfonts}
\usepackage{cite}
\usepackage{enumitem}
\usepackage{xcolor}
\usepackage{color}
\usepackage{mathtools}
\usepackage{caption} 
\usepackage{subcaption} 
\usepackage{euscript} 
\usepackage{float}
\usepackage{diagbox}
\usepackage[title]{appendix}
\usepackage{tabularx}

\renewcommand\d{{\rm d}}

\newcommand{\D}{{\cal D}}

\newcommand{\Pc}{\mathscr{P}}

\newcommand{\Fc}{\mathscr{F}}

\newcommand{\be}{\begin{equation}}
\newcommand{\ee}{\end{equation}}
\newcommand{\dis}{\displaystyle}
\renewcommand{\thefootnote}{\fnsymbol{footnote}}

\renewcommand{\O}{{\cal O}}
\newcommand{\cst}{{\rm const.}}

\newcommand{\ie}{{\em i.e.} }
\newcommand{\eg}{{\em e.g.} }

\renewcommand{\and}{\mbox{and}}

\newcommand{\esp}{\phantom{\!\!\overset{\displaystyle |}{|}}}

\newcommand{\bm}{\boldmath} 

\catcode`\@=11
\def\marginnote#1{}
\newcount\hour
\newcount\minute
\newtoks\amorpm
\hour=\time\divide\hour by60 \minute=\time{\multiply\hour by60
\global\advance\minute by-\hour}
\edef\standardtime{{\ifnum\hour<12 \global\amorpm={am}%
        \else\global\amorpm={pm}\advance\hour by-12 \fi
        \ifnum\hour=0 \hour=12 \fi
        \number\hour:\ifnum\minute<10 0\fi\number\minute\the\amorpm}}
\edef\militarytime{\number\hour:\ifnum\minute<10 0\fi\number\minute}
\def\draftlabel#1{{\@bsphack\if@filesw {\let\thepage\relax
   \xdef\@gtempa{\write\@auxout{\string
      \newlabel{#1}{{\@currentlabel}{\thepage}}}}}\@gtempa
   \if@nobreak \ifvmode\nobreak\fi\fi\fi\@esphack}
        \gdef\@eqnlabel{#1}}
\def\@eqnlabel{}
\def\@vacuum{}
\def\draftmarginnote#1{\marginpar{\raggedright\scriptsize\tt#1}}
\def\draft{\oddsidemargin -.2truein
        \def\@oddfoot{\sl preliminary draft \hfil
        \rm\thepage\hfil\sl\today\quad\militarytime}
        \let\@evenfoot\@oddfoot \overfullrule 3pt
        \let\label=\draftlabel
        \let\marginnote=\draftmarginnote
   \def\@eqnnum{(\theequation)\rlap{\kern\marginparsep\tt\@eqnlabel}%
\global\let\@eqnlabel\@vacuum}  }
\def\thebibliography#1{
\vskip 0.5cm \centerline{\bf \Large References}
\list{
[\arabic{enumi}]}{\settowidth\labelwidth{[#1]}
\leftmargin\labelwidth
\advance\leftmargin\labelsep
\usecounter{enumi}}
\def\newblock{\hskip .11em plus .33em minus .07em}
\sloppy\clubpenalty4000\widowpenalty4000
\sfcode`\.=1000\relax}

\renewcommand{\theequation}{\arabic{section}.\arabic{equation}}
\renewcommand{\section}{\setcounter{equation}{0}\@startsection
{section}{1}{0mm}{-\baselineskip}{0.5\baselineskip} {\normalfont\Large\bfseries}}
\renewcommand{\subsection}{\@startsection
{subsection}{2}{0mm}{-\baselineskip}{0.5\baselineskip} {\normalfont\large\bfseries}}
\renewcommand{\subsubsection}{\@startsection
{subsubsection}{3}{0mm}{-\baselineskip}{0.5\baselineskip}
{\normalfont\normalsize\slshape}}

\begin{document}


\begin{titlepage}
\begin{flushright}
CPHT-RR076.092021, October 2021
\vspace{0.0cm}
\end{flushright}
\begin{centering}
{\bm\bf \Large Probability distribution for the quantum universe \\ }

\vspace{10mm}

 {\bf Alex Kehagias,$^1$\footnote{kehagias@central.ntua.gr} {\bf Herv\'e Partouche$^2$}\footnote{herve.partouche@polytechnique.edu} and Nicolaos Toumbas$^3$\footnote{nick@ucy.ac.cy} 
 }

 \vspace{3mm}

$^1$  {\em Department of Physics, School of Natural Sciences, NTUA \\15780 Athens, Greece}

$^2$  {\em CPHT, CNRS, Ecole polytechnique, IP Paris, \\F-91128 Palaiseau, France}

$^3$ {\em Department of Physics, University of Cyprus, \\Nicosia 1678, Cyprus}

\end{centering}
\vspace{0.5cm}
$~$\\
\centerline{\bf\Large Abstract}\\
\vspace{-0.6cm}

\begin{quote}

We determine the inner product on the Hilbert space of wavefunctions of the universe by imposing the Hermiticity of the quantum Hamiltonian in the context of the minisuperspace model. The corresponding quantum probability density reproduces successfully the classical probability distribution in the $\hbar \to 0$ limit, for closed universes filled with a perfect fluid of index $w$. When $-1/3<w\le 1$, the wavefunction is normalizable and the quantum probability density becomes vanishingly small at the big bang/big crunch singularities, at least at the semiclassical level. Quantum expectation values of physical geometrical quantities, which diverge classically at the singularities, are shown to be finite. 

\end{quote}

\end{titlepage}
\newpage
\setcounter{footnote}{0}
\renewcommand{\thefootnote}{\arabic{footnote}}
 \setlength{\baselineskip}{.7cm} \setlength{\parskip}{.2cm}

\setcounter{section}{0}


\section{Introduction}

Quantum effects are essential in understanding the early-universe dynamics. In particular, it is important to determine the wavefunction of the universe, which is expected to specify the initial conditions for the cosmological evolution. A first attempt towards this direction has been undertaken by Wheeler and DeWitt who defined a Schr\"odinger-like equation, known as the Wheeler-DeWitt (WDW) equation \cite{DeWitt}, satisfied by such a wavefunction \cite{Vilenkin3,Vilenkin4,Linde,Linde2,Ruba, LindeBook}. Later, Hartle and Hawking~\cite{HH} introduced a wavefunction for a closed universe as a path integral satisfying the ``no-boundary proposal'' \cite{Halliwell,Brown,Turok1,Turok2, Halli-Hartle}. However, some issues arise from both viewpoints, which can be resolved \cite{Rpsi, Pro}. This can be illustrated in the simpler framework of  the minisuperspace model, where the closed universe is assumed to be homogeneous and isotropic with a positive cosmological constant, so that the scale factor is the only dynamical degree of freedom:  
  

\noindent $\bullet$ Field redefinitions $a=A(q)$ of the scale factor $a(t)$ leave the classical action invariant. However, in the path integral approach, the measures $\D a$ and $\D q$ are not equal, being related by a Jacobian. Hence, there exist infinitely many inequivalent definitions for the wavefunction in terms of path integrals of the form 
\be
\int{\D N\over {\rm Vol(Diff)}}\int \D q \,e^{iS[N,q]}\, , 
\ee
where $N$ is the lapse function, $\rm Vol(Diff)$ is the volume of the time-reparametrization group and $S$ is the gravitational action. This leads to different results, as can already be  seen at the semiclassical level \cite{Rpsi}.\footnote{\label{f1} To compute the path integrals, one replaces $\int\D N/{\rm Vol(Diff)}$ by $\int_0^{+\infty}\d \ell\times 1$, where $\ell$ is the invariant length of the time interval, and the factor $1$ stands for a Fadeev-Popov determinant, which turns out to be trivial. As explained in Sect.~5.3 of Ref.~\cite{Rpsi}, previous work often employs a gauge for the lapse function that depends on the scale factor, $N(t)=\ell/a(t)$ \cite{Halliwell,Turok1, Halli-Hartle}. In that case, the limits of the time integral in the action $S$ depend on the trajectory $q(t)$. Therefore, it is not valid to only vary the Lagrangian with respect to $q$ to find the saddle points of the action. The authors proceed to treat the limits of integration in the action as constants, but this changes the classical problem to be quantized. The motivation for the gauge $N(t)=\ell/a(t)$ is that the field redefinition $a=A(q)=\sqrt{q}$ renders the Lagrangian quadratic in $q$. This then would seem to turn the path integral into a Gaussian one. However one should keep track of the change in the limits of integration. Indeed, the computation at the semiclassical level can be done in the gauge $N(t)=\ell$, for which the limits of integration are independent of the trajectory~\cite{Rpsi}, yielding different results from \cite{Halliwell,Turok1, Halli-Hartle}.}
 
\noindent $\bullet$ Each choice of field $q(t)$ yields a different classical Hamiltonian ${\cal H}$, which vanishes. At the quantum level, the constraints ${\cal H}=0$ translate into different WDW equations accompanied with additional ambiguities in their forms, due to the non-commutativity of the canonical variable $q$ and its conjugate momentum $\pi_q$ \cite{DeWitt}. The ambiguities can be partially implemented by replacing
\be
\pi_q^2 = {1\over \rho(q)}\,\pi_q\,\rho(q)\,\pi_q~~\longrightarrow ~~ -{i\over \rho(q)}{\partial\over \partial q}\Big(\!-i\rho(q){\partial\over \partial q}\Big)\, .
\ee
 
\noindent $\bullet$ By imposing that the path-integral wavefunctions are solutions of the WDW equations, one obtains that \cite{Rpsi}\footnote{This resolution of the ambiguity in form of the WDW equation differs from that derived in Ref.~\cite{Halliwell}, which presents the issue raised in Footnote~\ref{f1}.}
 \be
\rho(q)=A(q)^{-{3\over 4}}|A'(q)|^{-{3\over 2}}\, .
 \ee  
  
\noindent $\bullet$ For each prescription, one defines an inner product on the Hilbert space,
\be
\langle \Psi_1|\Psi_2\rangle = \int \d q \, \mu(q) \, \Psi_1(q)^*\Psi_2(q)\, , 
\ee 
such that the corresponding Hamiltonian ${\cal H}$ is Hermitian. This requirement determines the measure $\mu(q)$ on the Hilbert space to be \cite{Rpsi}
\be
\mu(q) = A(q)\, A'(q)^2\, \rho(q)\, .
\ee

\noindent $\bullet$ The key point is that at the semiclassical level, one finds
\be
\Psi(q)\propto {\sqrt{|A'(q)|\over \mu(q)}}\, , 
\ee
which implies that the inner product is independent of the prescription, at this level of approximation. This strongly suggests that the various choices of fields $q$ do not yield different quantum gravity theories with the same classical limit. Hence, we may choose $q\equiv a$, which leads to
 \be
 \mu(a) = a\, \rho(a)\, .
 \label{arho}
 \ee
 
\noindent $\bullet$ Notice that in the literature, a widely used inner product on the Hilbert space is based on the measure $\mu_{\rm HH}(a)=\rho(a)$ employed by Hartle and Hawking in Ref. \cite{HH}. Their choice follows by imposing Hermiticity of the Wheeler-DeWitt operator $a{\cal H}$ rather than ${\cal H}$.  


The relation $\mu(a)=a \rho(a)$ arising from the Hermiticity of the quantum Hamiltonian is valid in the more general context where the closed, homogeneous and isotropic universe is filled with a perfect fluid of state equation $p_m=w\rho_m$. In this work, we test the consistency of this measure by showing that the quantum probability density reproduces in the $\hbar\to 0$ limit the classical probability distribution. In the context of quantum cosmology see {\it e.g.} Refs. \cite{HP, Vilenkin5, Halliwell2,HGC}. To find the classical probability distribution, we compute the ratio of the time the scale factor  lies in the infinitesimal range $[a,a+\d a]$ to the total duration of the cosmological evolution. The total duration is finite when the index $w$ of the perfect fluid satisfies $w>-1/3$. If $-1\le w \le -1/3$, the duration is infinite but relative probabilities can be defined, leading again to a consistency between the classical and quantum pictures. 

Notice that when $w>-1/3$, an initial singularity exists at the classical level. However, the fact that the wavefunction is normalizable implies that the classical singularity is innocuous in the quantum case. This is similar to the problem of the hydrogen atom, where the classical singularity at the center of the atom is resolved quantum mechanically.\footnote{For $w=-1/3$, the wavefunction is non-normalizable but $\int_0^\alpha \d a \mu |\Psi|^2$ is finite for any $\alpha>0$, implying the classical initial singularity to be again innocuous. }

Indeed, to obtain the measure $\mu$ for more general Hamiltonians in higher $d$, one could generalize the derivation in the Appendix of Ref.~[14], see Eq. (A.21) - Eq. (A.25). This derivation is partially based on Section 9 of DeWitt's paper [1], where he obtains Eqs. (9.3) and (9.4) \footnote{We mention that $i$ factors in the r.h.s. of these equations are missing.}, valid for more general Hamiltonians in arbitrary $d$. In fact, one has to include in all steps of DeWitt's proof a measure $\mu$ and $\rho$-like functions (arising from operator ordering ambiguities). One could then generalize Eq. (A.25) of Ref.~[14] for a more general Hamiltonian in arbitrary $d$. To determine $\mu$ from this result, one must follow the procedure below Eq. (A.25) of Ref.~[14]: Since the functions $\Psi_1$ and $\Psi_2$ are arbitrary, the form of the measure $\mu$ is extremely constrained. So one obtains the analogous result given below Eq. (A.27) of Ref.~[14].

The paper is organized as follows. In Sect. \ref{class}, we compute the classical probability distribution in the case of a closed universe filled with a perfect fluid with arbitrary index $w\in[-1,1]$. In Sect.~\ref{micro}, we derive the WDW equation from a microscopic description of the fluid based on a scalar field. In Sect. \ref{quant}, we obtain the quantum probability density at the semiclassical level and find perfect agreement with the classical probability distribution in the $\hbar\to 0$ limit. A summary of our results  can be found in the conclusion, Sect. \ref{cl}. The Hermiticity of the Hamiltonian is also important in other approaches. In Ref.~\cite{DiGioia}, the authors discuss the problem of time and non-unitarity emerging in an adiabatic, WKB approach concerning quantum gravity corrections to quantum field theory.

\section{Classical probability distribution}
\label{class}
In this section, we consider classical cosmological evolutions in order to derive the associated probability distributions.

\subsection{Classical cosmological evolution}

Our starting point is gravity coupled to matter with action
\be
{\cal S}=\int \d^4 x \sqrt{-g}  \left(\frac{R}{2} +{\cal L}_m\right),
\ee
where ${\cal L}_m$ is the matter Lagrangian density. We work in units $8\pi G=M_{\rm P}^2=1$. The equations of motion are 
\be
 R^{\mu\nu}-\frac{1}{2} g^{\mu\nu} R=T^{\mu\nu}\,,
 \ee
 where $T^{\mu\nu}$ is the matter energy-momentum tensor.
We will examine the case where $T^{\mu\nu}$  describes a perfect fluid with equation of state $p_m=w \rho_m$, where $\rho_m$ and $p_m$ are the energy density and pressure, while the index $w$  is in the range $-1\le w \le 1$. For FRW metric
 \be
  \d s^2=-\d t^2+a^2(t) \d\Omega_3^2\, ,
  \ee
  where $\d\Omega_3^2$ is the line element of the unit 3-sphere, the Friedmann equations are 
  \begin{align}
  3H^2&=\rho_m-\frac{3}{a^2}\,, \label{1st} \\
  3\,\frac{\ddot a}{a}&=-{1\over 2}\, (\rho_m+3P_m)=-{1\over 2}\,(1+3w)\rho_m\, , \label{2nd}
  \end{align}
with $H=\dot a /a$ denoting the Hubble parameter.
  Conservation of energy-momentum specifies the energy density to be of form 
\be
\rho_m=\left\{\begin{array}{lll}
\displaystyle {3\, a_0^{3w+1}\over a^{3(w+1)}}\, , &a_0>0\, , &\mbox{for}\quad  \displaystyle w\neq -{1\over 3}\, ,\\
\displaystyle{3\, V_0\over a^2}\, , &V_0>1\, , & \mbox{for} \quad \displaystyle w= -{1\over 3}\, .\esp
\end{array}\right.
 \label{rm}
\ee

With this notation and for $w\neq -1/3$, the first Friedmann equation can be written as 
\be
\dot a^2=\left({a_0\over a}\right)^{3w+1}-1\, ,
\label{fri2}
\ee
which shows that there is always a turning point $(\dot a=0$) when  $a=a_0$. Two cases must be distinguished:

\noindent {\bf\bm $\bullet$  $w>-1/3$: }\\
In this case, the scale factor satisfies $a\leq a_0$. The universe is decelerating, as follows from Eq. (\ref{2nd}), and $\dot a\to \pm \infty$ when $a\to 0$. Therefore, the evolution starts/ends with a big bang/big crunch, and the scale factor is maximal when $a=a_0$.

\noindent {\bf \bm $\bullet$  $w<-1/3$: }\\
In this case, the scale factor satisfies $a\geq a_0$. The trajectory is accelerating and $\dot a\to \mp \infty$ when $a\to +\infty$. Thus, the evolution describes a bouncing universe, with minimal size when $a=a_0$. There is a contracting/expanding phase before/after the bounce, from/to $a=+\infty$.

In order to obtain the explicit solutions for $w\neq -1/3$, we use Eq. (\ref{fri2})  in the form
\be
{a^{3w+1\over 2}\, \d a\over \sqrt{a_0^{3w+1}-a^{3w+1}}}=\pm \d t \, .
\ee
The l.h.s. is always integrable at $a=a_0$. Upon integration, one sees that the lifetime of the universe is  finite for $w>-1/3$ and infinite for $-1\le w<-1/3$.\footnote{This will be related to the fact  that the quantum wavefunction will be normalizable for $w>-1/3$ and non-normalizable for $-1\le w< -1/3$.} Time as a function of the scale factor is given by 
\begin{align}
\pm t&=a_0\int_1^{a\over a_0}{x^{3w+1\over 2}\, \d x\over \sqrt{1-x^{3w+1}}}\label{sol} \\
&={2a_0\over 3(w+1)}\,\Big({a\over a_0}\Big)^{{3\over 2}(w+1)}{}_2F_1\bigg({1\over 2}, {1\over 2}+{1\over 3w+1},{3\over 2}+{1\over 3w+1}; \Big({a\over a_0}\Big)^{3w+1}\bigg)-a_0 \,\sqrt{\pi}\, {\Gamma\Big({1\over 2}+{1\over 3w+1}\Big)\over \Gamma\Big({1\over 3w+1}\Big)} \, , \nonumber 
\end{align}
where we have chosen the time origin such that the cosmological evolution is symmetric under time reversal $t\to -t$. Notice that in the last line, one may think that there is a pathology when the Gamma function in the numerator diverges, {\it i.e.} when $1/2+1/(3w+1)=-n$ where $n\in\mathbb{N}$; that is when $w=-1/3-2/(6n+3)$, which lies in the range $[-1,-1/3)$. However, this is only an artefact of this way of writing things, since the first line is perfectly well behaved for any $w\neq -1/3$.

Finally, for the remaining case $w=-1/3$, one finds two possible cosmological evolutions described by the ``$+$'' or ``$-$'' branches of the following equation  
\be
a = \pm \sqrt{V_0-1}\; t\, .
\ee

\subsection{Classical probability distributions for \bm $w>-1/3$}

The finite lifetime of the universe, as follows from Eq. (\ref{sol}), is 
\begin{align}
\Delta t &= 
-2a_0\int_1^0{x^{3w+1\over 2}\, \d x\over \sqrt{1-x^{3w+1}}}\, \nonumber \\
&= a_0 \,2\sqrt{\pi}\, {\Gamma\big({1\over 2}+{1\over 3w+1}\big)\over \left|\Gamma\big({1\over 3w+1}\big)\right|}\, . \esp
\end{align}
It satisfies $\Delta t\to +\infty$ when $w\to (-1/3)_+$.

In order to compute the classical probability distribution, we must find the time the scale factor spends in the interval $[a,a+\d a]$. This time turns out to be 
\be
\delta t=2|\d t| = {2a^{3w+1\over 2}\,  \d a\over \sqrt{a_0^{3w+1}-a^{3w+1}}}\, , 
\ee
where the factor of $2$ takes into account the contracting and expanding phases. Then, 
the classical probability distribution as a function of $a$ times $\d a$ is 
\begin{align}
P_{cl}(a)\,\d a &={\delta t\over \Delta t} \nonumber \\
&={1\over \sqrt{\pi}}\, {\left|\Gamma\big({1\over 3w+1}\big)\right|\over \Gamma\big({1\over 2}+{1\over 3w+1}\big)}\, {a^{3w+1\over 2}\,  \d a\over a_0\sqrt{a_0^{3w+1}-a^{3w+1}}}\, ,
\label{pa}
\end{align}
where $a\le a_0$.
By construction, the total probability is normalized to unity,
\begin{equation}
\int_0^{a_0} P_{cl}(a)\,\d a =1. 
\end{equation}
Notice that the classical probability distribution is infinite at the turning point $a=a_0$, where the rate of expansion is zero, and vanishes at $a=0$ where the rate of expansion/contraction is infinite.

\subsection{Classical probability distributions for \bm $-1\le w\le -1/3$}

In this case, the lifetime of the universe is infinite but nevertheless we can consider ``relative probabilities.'' To this end, let us define the quantity 
\be
P_{cl}(a)\, \d a =\delta t=\left\{\begin{array}{ll}2|\d t| = \displaystyle {2a^{3w+1\over 2}\,  \d a\over \sqrt{a_0^{3w+1}-a^{3w+1}}}\, , &\mbox{for}\quad -1\le \displaystyle w<-{1\over 3}\, ,\\
\displaystyle |\d t| = {\d a\over \sqrt{V_0-1}}\, , \esp& \mbox{for}\quad \displaystyle w=-{1\over 3}\, ,
\end{array}
\right.
\label{pcl}
\ee
which cannot be interpreted as an absolute probability distribution since its integral over $a\in[a_0,+\infty)$ or $a\in[0,+\infty)$ is infinite. However, ratios of such quantities (at different values of the scale factor) can be understood as relative probabilities.  


\section{Microscopic derivation of WDW equation for any state equation}
\label{micro}

 In this section, our aim is to derive the WDW equation in minisuperspace, when the closed universe is filled with a perfect fluid of index $w\neq-1,0$.\footnote{The case $w=0$ can be studied as a limiting case, taking $w$ to be a very small positive constant. The case $w=-1$ can be easily derived by adding a positive cosmological constant and no matter field, see \eg  \cite{Rpsi}.}

Let us consider the microscopic model involving a derivatively coupled scalar field, with action \cite{Mukhanov, Brand}  
\begin{equation}
{\cal S}=\int_M d^4 x\sqrt{-g}\left\{\frac{R}{2}+\mbox{sign}(w)\left(-\frac{1}{2} \partial_\mu \phi 
\partial^\mu \phi\right)^{\frac{w+1}{2w}}\right\}+{\cal{S}}_{\rm boundary} \, ,\label{action0}
\end{equation}
where $M$ is spacetime and 
\begin{equation}
{\cal{S}}_{\rm boundary}=-\int_{\partial M} d^3 y \sqrt{h}\, K
\ee
is the Gibbons-Hawking-York boundary term. In the above expression, $\partial M$ is a spacelike boundary of extrinsinc curvature $K$ and induced metric $h_{ij}$. The energy-momentum tensor for the scalar field is given by 
\begin{equation}
T^{\mu\nu}=g^{\mu\nu}\,\mbox{sign}(w) \left(-\frac{1}{2} \partial_\kappa \phi 
\partial^\kappa \phi\right)^{\frac{w+1}{2w}}+\frac{w+1}{2|w|} \left(-\frac{1}{2} \partial_\kappa \phi 
\partial^\kappa \phi\right)^{\frac{1-w}{2w}}\partial^\mu  \phi \partial^\nu \phi\,.
\end{equation}
It turns out to be of  the form of the energy-momentum tensor of an irrotational fluid,
\begin{equation}
T^{\mu\nu}=(\rho_m+p_m)\,u^\mu u^\nu+p_m\,g^{\mu\nu} \, ,
\end{equation}
where 
\begin{equation}
p_m=\mbox{sign}(w)\left(-\frac{1}{2} \partial_\kappa \phi 
\partial^\kappa \phi\right)^{\frac{w+1}{2w}}\,, \qquad u^\mu=\frac{\partial^\mu \phi}{\sqrt{- \partial_\kappa \phi 
\partial^\kappa \phi }}\,,  
\end{equation}
and equation of state 
\begin{equation}
p_m=w\rho_m\,.  \label{eqst}
\end{equation}

In minisuperspace, the $4$-metric and the scalar field are of the form 
\begin{equation}
\d s^2=-N(x^0)^2 \big(\d x^0\big)^2+a^2(x^0) \d\Omega_3^2\, ,\qquad\phi=\phi(x^0)\,,
\end{equation}
and the action becomes 
\begin{equation}
{\cal S}=v_3 \int dx^0 \left( -3\, \frac{a \dot{a}^2}{N}+3 N a +\frac{\mbox{sign}(w)}{2^{\frac{w+1}{2w}}}\, \frac{a^3}{N^{\frac{1}{w}}}\, |\dot{\phi}|^{\frac{w+1}{w}} 
\right),
\end{equation}
where $v_3=2\pi^2$ is the volume of the unit $3$-sphere.  Denoting $L$ the Lagrangian,  the generalized momenta are given by 
\begin{eqnarray}
&&\pi_a=\frac{\partial L}{\partial \dot a }=-6 v_3\, \frac{a}{N}\, \dot a\, , \qquad  \pi_N={\partial L\over \partial \dot N}=0\, , \nonumber\\
&&\pi_\phi=\frac{\partial L}{\partial \dot \phi }= \frac{v_3}{2^{\frac{w+1}{2w}}} \,\frac{a^3}{N^{\frac{1}{w}}}\, \frac{w+1}{|w|} \,
|\dot \phi|^{\frac{1}{w}} \,\, \mbox{sign}(\dot \phi)\, .
\end{eqnarray}
The classical Hamiltonian can then be expressed as 
\begin{align}
\frac{{\cal H}}{N}&=-\frac{1}{12 v_3}\,\frac{\pi_a^2}{a}-3 v_3  a+\frac{|w|^w}{(w+1)^{w+1}}\, \frac{2^{\frac{w+1}{2}}}{v_3^w }\, \big(\mbox{sign}(\pi_\phi)\big)^{w+1}\, \frac{\pi_\phi^{w+1}}{a^{3w}}\nonumber \\
&=-\frac{1}{12 v_3}\,\frac{\pi_a^2}{a}-3 v_3  a+C \, \frac{|\pi_\phi|^{w+1}}{a^{3w}}\,,
\end{align}
where the constant $C$ is 
\begin{equation}
C= \frac{|w|^w}{(w+1)^{w+1}}\, \frac{2^{\frac{w+1}{2}}}{v_3^w } \,.
\end{equation}
It satisfies
\begin{equation}
\frac{{\cal H}}{N}=-\frac{\partial L}{\partial N}=0\, ,
\end{equation}
by the equation of motion of the lapse $N$, which takes the form 
\begin{equation}
{3\over N^2}\, H^2=\rho_m-\frac{3}{a^2}\, ,
\end{equation}
where
\begin{equation}
\rho_m={C\over v_3} \, \frac{|\pi_\phi|^{w+1}}{a^{3(w+1)}}\ge 0\,. \label{rm1}
\end{equation}
Note that $\pi_\phi=\cst$, as follows from the Hamilton equation $\dot \pi_\phi=-\partial {\cal H}/\partial \phi=0$. As a result, the scaling of the energy density $\rho_m$ is the expected one for a perfect fluid with equation of state 
(\ref{eqst}). Comparing Eqs. (\ref{rm}) and (\ref{rm1}), we find that 
\begin{equation}
{C\over v_3} \,|\pi_\phi|^{w+1}=\left\{\begin{array}{ll} 3 a_0^{3w+1}\, ,&\mbox{for}\quad \dis w\neq -\frac{1}{3}\,,\\
3V_0\, ,&\dis \mbox{for}\quad w=-\frac{1}{3}\,.\esp
\end{array}
\right.
\end{equation}

To canonically quantize the theory, we replace 
\begin{equation}
\pi_a~  \longrightarrow~ -i \hbar \,\frac{\partial }{\partial a}\, , \qquad \pi_\phi~\longrightarrow~- i \hbar \,\frac{\partial }{\partial \phi}\,. 
\end{equation}
However, because classically 
\be
{\pi_a^2\over a} = {1\over a\, \rho_1\,\rho_2}\, \pi_a \,\rho_1\, \pi_a\,\rho_2 \, , 
\ee
where $\rho_1(a)$, $\rho_2(a)$ are arbitrary functions of the scale factor, the process of quantization yields an ambiguity in the form of the WDW equation of the wavefunction $\Phi(a,\phi)$,
\be
{{\cal H}\over N}\,\Phi(a,\phi)\equiv {\hbar^2\over 12v_3}\, {1\over a \rho_1\rho_2}\, {\partial\over \partial a}\Big[\rho_1\, {\partial\over \partial a}\big(\rho_2\Phi\big)\Big]-
3 v_3  a \,\Phi+\frac{C}{a^{3w}}\left|-i \hbar \,\frac{\partial \Phi}{\partial \phi}\right|^{w+1}=0\, .
\ee
 Redefining 
\be
\rho= \rho_1\rho_2^2\, , \qquad \omega={(\rho_1\rho_2')'\over \rho_1\rho_2}\, , 
\ee
one obtains the more convenient form 
\be
{{\cal H}\over N}\,\Phi(a,\phi)\equiv \left\{\frac{\hbar^2}{12 v_3a}\left(\frac{1}{\rho}\,\frac{\partial }{\partial a}\Big(\rho\,\frac{\partial}{\partial a}\Big)+\omega\right)-
3 v_3  a +\frac{C}{a^{3w}}\left|-i \hbar\, \frac{\partial }{\partial \phi}\right|^{w+1}\right\}\Phi(a,\phi)=0\, ,
\label{wdw2}
\ee
where $\rho$ and $\omega$ depend on $a$. We will see in the next section that none of the functions $\rho$ and $\omega$ need to be known (at least) at the semiclassical level~\cite{Rpsi}, and to recover the classical probability distribution from the quantum picture. Moreover, notice that  there is no issue of non-commutativity for the term $\pi_\phi^{w+1}$ since the canonical variable $\phi$ does not appear in the Hamiltonian. Since Eq.~(\ref{wdw2}) does not depend on $\phi$,  the wave function $\Phi(a,\phi)$ can be expressed as a  Fourier mode  
\be
\Phi(a,\phi)=\Psi(a)\,  e^{i\frac{\lambda}{\hbar}\phi}\, ,
\ee
where $\lambda$ is an arbitrary real constant. In that case,  one has $\pi_\phi=\lambda$ and  $\Psi(a)$  satisfies
\begin{equation}
\Psi''+\frac{\rho'}{\rho}\, \Psi'+\omega\,\Psi=\frac{\Pc(a)}{\hbar^2}\, \Psi\, ,
\label{wdw1}
\end{equation}
where 
\begin{align}
\Pc(a)&=36 v_3^2 \,a^2\left(1-\frac{C|\lambda|^{w+1}}{3v_3 a^{3w+1}}\right)\nonumber \\
&=\left\{\begin{array}{ll}36 v_3^2 \,a^2\displaystyle \left(1-\Big({a_0\over a}\Big)^{3w+1}\right),&\mbox{for}\quad w\neq -1/3\,,\esp\\
36 v_3^2 \,a^2\left(1-V_0\right),&\mbox{for}\quad  w= -1/3\,.\esp
\end{array}\right.\label{p}
\end{align}
Finally, as shown in Ref. \cite{Rpsi}, the quantum Hamiltonian for any fixed $\lambda$  is Hermitian with respect to the inner product
\begin{equation}
  \langle \Psi_1|\Psi_2\rangle=\int_0^{+\infty} \d a \, \mu(a)\, \Psi_1(a)^*\Psi_2(a)\, ,
  \end{equation}  
where the measure $\mu(a)$ of the Hilbert space is given in Eq. (\ref{arho}). 

\section{Quantum probability density}
\label{quant}

Our goal is to derive the quantum probability distributions and show their consistencies with the classical results.

Since we are interested in the classical limit, it is enough to solve the WDW equation (\ref{wdw1}) at the semiclassical level using the WKB method. To this end, we define
\be
\Psi = \exp\!\Big[ {i\over \hbar} \Big( \Fc_0 + {\hbar\over i}\Fc_1 + \O(\hbar^2) \Big) \Big] \, ,
\ee
which can be used in Eq.~(\ref{wdw1}) to get
\begin{align}
\left\{\!\!\begin{array}{rl}
\Fc_0'^2 \!\!\!\!\!&= -\Pc\, , \\
\Fc_1' \!\!\!\!\!&=\displaystyle -{1\over 2}\Big({\Fc_0''\over \Fc_0'}+{\rho'\over \rho}\Big)
\end{array}
\right. \Longrightarrow\quad \Fc_1 = \ln \!\Big(|\Pc|^{-{1\over 4}}\rho^{-{1\over 2}}\Big)+{\cst}\, .
\end{align}
As anticipated before, one sees that the function $\omega$ plays no role at the semiclassical level. Moreover, the expression of $\Fc_1$ implies $\Psi\propto 1/\sqrt\rho$, so that  the probability density $\mu |\Psi|^2$, where $\mu = a \rho$, is independent of the function $\rho$, at least at the semiclassical level.

In the quantum case, the scale factor $a$ can take any value from $0$ to $+\infty$.  When $w\neq -1/3$,  the expression of the probability density then depends on whether $a$ is greater or smaller than $a_0$:  

\noindent $\bullet$ {\bm\bf  $a<a_0$ when $w>-1/3$, or $a>a_0$ when $w<-1/3$:}\\
In this case,  $\Pc(a)<0$ so that 
\be
\Fc_0'=\epsilon\,  \sqrt{-\Pc}\quad \Longrightarrow\quad \Fc_0=\epsilon \, {4v_3\over 1-w} \, a^2  \Big({a_0\over a}\Big)^{3w+1\over 2}\Big[1+O\Big(\Big({a\over a_0}\Big)^{3w+1}\Big)\Big] +{\cst}\, ,
\label{e1}
\ee
where $\epsilon = \pm 1$ and the $O$-term is independent of $\epsilon$. As a result, the wavefunction takes the form
\be
\Psi (a)= \sum_\epsilon N_\epsilon(a_0) \,{\displaystyle \exp\!\Big[\epsilon \, {i\over \hbar}\, {4v_3\over 1-w}  \,a^2  \Big({a_0\over a}\Big)^{3w+1\over 2}\big[1+O\big(({a/ a_0})^{3w+1}\big)\big]\Big]\over \sqrt{\rho} \; a^{1-3w\over 4}\big(a_0^{3w+1}-a^{3w+1}\big)^{1\over 4}}\, \big(1+O(\hbar)\big)\, ,
\ee
where $N_+(a_0)$, $N_-(a_0)$ are integration constants that depend on $a_0$ but not on $\hbar$. 
%
Note that for any scale factor in the range  $0<a<a_0$, the WKB approximation can be trusted.  At $a=0$ and $a=a_0$, the adiabaticity condition is not satisfied and  one has to solve the full WDW equation around these points and match the solutions to the WKB one in the usual way, see \eg Ref.~\cite{wkb}. 

The quantum probability density, using the measure in Eq. (\ref{arho}), is thus 
\begin{align}
P(a) &=a\rho|\Psi|^2\nonumber \\
&={a^{3w+1\over 2}\over \sqrt{a_0^{3w+1}-a^{3w+1}}}\, \big(1+O(\hbar))  \bigg\{|N_+|^2+|N_-|^2\nonumber\\
&~~~~~~~~~~~~~~~+ N_+\bar N_-  \exp\!\Big[{i\over \hbar}\, {8v_3\over 1-w}  \,a^2  \Big({a_0\over a}\Big)^{3w+1\over 2}\big[1+O\big(({a/ a_0})^{3w+1}\big)\big]\Big]+\mbox{c.c.} \bigg\} \, .
\end{align}
However, in the  $\hbar \to 0$ limit, the frequency of the oscillatory terms blows up and the latter can be omitted. 
As a result, we obtain
\be
P(a)\simeq\Big(|N_+(a_0)|^2+|N_-(a_0)|^2\Big) {a^{3w+1\over 2}\over \sqrt{a_0^{3w+1}-a^{3w+1}}}\, \big(1+O(\hbar)\big) \, .
\label{e2}
\ee
Therefore, $P$ and $P_{cl}$ (see Eqs. (\ref {pa}) and (\ref{pcl})) become proportional in the $\hbar\to 0$ limit.  

\noindent $\bullet$ { \bm \bf $a>a_0$ when $w>-1/3$, or $a<a_0$ when $w<-1/3$:}\\
This is the classically forbidden region. In this case, we have $\Pc(a)>0$ so that 
\be
\Fc_0'=\epsilon \, i \sqrt{\Pc}\quad \Longrightarrow\quad \Fc_0=\epsilon \, i\, 3v_3 \,a^2 \Big[1+O\Big(\Big({a_0\over a}\Big)^{3w+1}\Big)\Big] +\cst\, ,
\label{e3}
\ee
where $\epsilon = \pm 1$ and again the $O$-term are independent of $\epsilon$. The wavefunction now takes the form
\be
\Psi (a)= \sum_\epsilon M_\epsilon(a_0) \,{\displaystyle \exp\!\Big[\!- \! {\epsilon\over \hbar} \,3v_3\,a^2  \big[1+O\big(({a_0/ a})^{3w+1}\big)\big]\Big]\over \sqrt{\rho} \, a^{1-3w\over 4}\big(a^{3w+1}-a_0^{3w+1}\big)^{1\over 4}}\, \big(1+O(\hbar)\big)\, , 
\ee
where $M_+(a_0)$, $M_-(a_0)$ are integration constants independent of $\hbar$. 

The quantum probability density is 
\begin{align}
P(a)&=a\rho|\Psi|^2\nonumber \\
&={a^{3w+1\over 2}\over \sqrt{a^{3w+1}-a_0^{3w+1}}}\, \big(1+O(\hbar))   \bigg\{|M_-|^2 \exp\!\Big[{6\over \hbar}\, v_3 \,a^2 \big[1+O\big(({a_0/ a})^{3w+1}\big)\big]\Big]\nonumber\\
&~~~~~~~~~~~~+M_+\bar M_-+\bar M_+M_-+|M_+|^2 \exp\!\Big[\!-\!{6\over \hbar}\, v_3 \,a^2 \big[1+O\big(({a_0/ a})^{3w+1}\big)\big]\Big] \bigg\}\, .
\label{e4}
\end{align}
Considering for $w>-1/3$ a normalizable state,  we have 
\be
1=\int_0^{a_0} a\rho|\Psi|^2\d a+\int_{a_0}^{+\infty} a\rho|\Psi|^2\d a\, ,
\ee
where the first term in the r.h.s. is always finite whereas the second term is finite only when $M_-=0$. More generally, for $w\neq -1/3$, imposing that a finite  $\hbar\to 0$ limit exists in Eq. (\ref{e4}), we conclude that $M_-=0$.   In this case, we obtain 
\be
P(a)=|M_+(a_0)|^2 \, {a^{3w+1\over 2}\over \sqrt{a^{3w+1}-a_0^{3w+1}}}\,  \exp\!\Big[\!-\!{6\over \hbar}\, v_3 \,a^2 \big[1+O\big(({a_0/ a})^{3w+1}\big)\big]\Big]\big(1+O(\hbar)\big)\, .
\label{e5}
\ee
As a result, in the classical limit, the quantum probability density in the classically excluded region is exponentially suppressed. As expected, we have $P(a)\to 0$ when $\hbar\to 0$.

\noindent $\bullet$ {\bf \bm   $w=-1/3$:}\\
In this particular case, we have $\Pc(a)<0$ for all $a\ge 0$ and we find
\be
\Psi (a)= \sum_\epsilon N_\epsilon(V_0) \,{\displaystyle \exp\!\Big[\epsilon \, {i\over \hbar}\, 3v_3  \sqrt{V_0-1} \;a^2\Big] \over \sqrt{a \rho} }\, \big(1+O(\hbar)\big)\, ,
\ee
where $\epsilon=\pm 1$.
Therefore, the quantum probability is 
\begin{align}
P(a)&=a\rho|\Psi|^2\nonumber \\
&= \, \big(1+O(\hbar))  \bigg\{|N_+|^2+|N_-|^2+ N_+\bar N_-  \exp\!\Big[{i\over \hbar}\, 6v_3 \sqrt{V_0-1}\;a^2 \Big]+\mbox{c.c.} \bigg\} \, .
\end{align}
The oscillatory terms can be omitted in the  $\hbar\to 0$ limit, so that 
\be
P(a) \simeq \big(|N_+|^2+|N_-|^2\big)\, \big(1+O(\hbar))  \,,
\ee 
which becomes proportional to $P_{cl}$ (see Eq. (\ref{pcl})). 


Summarizing, for any $w\in[-1,1]$ and any $a\ge 0$, the quantum probability density $P$ in the $\hbar\to 0$ limit is proportional to the classical one $P_{cl}$. 
 
 For $w>-1/3$, these densities are normalized to 1 and thus, comparing Eqs. (\ref{e2}) and (\ref{pa}), we can identify\footnote{One could relate $N_+$ and $N_-$ to $M_+$ by using the gluing relations between the two phases $a>a_0$ and $a<a_0$, using Airy functions which solve the WDW equation when $a\simeq a_0$ \cite{wkb}. In that case we would identify $N_+, N_-, M_+$.}
\be
|N_+(a_0)|^2+|N_-(a_0)|^2 = {1\over \sqrt{\pi}}\, {\left|\Gamma\big({1\over 3w+1}\big)\right|\over \Gamma\big({1\over 2}+{1\over 3w+1}\big)}\, {1\over a_0}\, .
\ee 
Moreover, it is important to stress that although there exists an initial singularity at the classical level, the fact that the wavefunction (at finite $\hbar$) is normalizable suggests that the classical singularity at $a\to 0$ is harmless. In fact, the semiclassical probability density satisfies
\be
P(a)\underset{a\to 0}\sim {1\over \sqrt{\pi}}\, {\left|\Gamma\big({1\over 3w+1}\big)\right|\over \Gamma\big({1\over 2}+{1\over 3w+1}\big)}\, {1\over a_0}\Big({a\over a_0}\Big)^{3w+1\over 2}\,  \big(1+O(\hbar)\big) \,,
\ee
which vanishes at $a=0$ exactly as in the case of the probability density for the position of the electron at the center of   the hydrogen atom.

We may proceed to calculate at the semiclassical level the expectation value of the extrinsic curvature $K$ associated with the spatial 3-sphere. The extrinsic curvature is equal to $6/a^2$, yielding 
\be
\langle K \rangle = \int_0^{+\infty} 6\, {P(a)\over a^2}\, \d a \,.
\ee
At the semiclassical level, in the neighborhood of $a=0$, the integrand is proportional to $a^{3(w-1)/2}$, showing that the contribution to the expectation value from the small $a$ region is finite, provided that 
\be
w> {1\over 3}\, .
\ee
Classically, we expect the very early universe to be dominated by the kinetic energy of moduli fields, \ie a perfect fluid of index $w=1$. Thus, although this implies a curvature singularity at the classical level, this is resolved at the quantum level since the expectation value of the extrinsic curvature is finite. Notice that this would not be the case had we used the Hartle-Hawking measure $\mu_{\rm HH}$ for the inner product of the Hilbert space.   

For  $-1\le w\le-1/3$, we have chosen an arbitrary normalization of $P_{cl}$ in Eq.~(\ref{pcl}) (since only ratios yielding relative probabilities make sense). Hence, we conclude that in the $\hbar\to 0$ limit, the quantum system based on non-normalizable wavefunctions reproduces consistently the classical relative probabilities. 

Notice that for $w\neq-1/3$, $P(a)$ becomes large when the scale factor is of the order of $a_0$; that is, when the classical closed universe is at its turning point and the energy density is extremal. The favored size of the universe ``created from nothing'' is thus of the order of $a_0$, for which the energy density is extremal.

Our approach can be readily generalized to higher dimensional cases when the Wheeler-DeWitt equation is a separable equation.
For example, we may add, as in Ref.~\cite{HH}, a conformally coupled scalar field $\phi$. Upon suitable rescalings ($\phi \to \chi/a$), the Hamiltonian (over the lapse function $N$) splits into a sum of two factors, ${\cal{H}}_1(a, \rho(a)) + {\cal{H}}_2(\chi)/a$, where ${\cal{H}}_1(a, \rho(a))$ is the Hamiltonian for the one-dimensional model we consider in the present work, and ${\cal{H}}_2(\chi)$ is a harmonic oscillator potential. Imposing the Hermiticity of the full Hamiltonian yields a measure $\mu(a,\phi)=\mu(a, \rho)\times 1$, where the factor $1$ stands for the usual ``trivial''  quantum mechanical measure on the Hilbert space spanned by the eigenfunctions of ${\cal{H}}_2(\chi)$, while $\mu(a,\rho)$ is the measure in Eq.~(\ref{arho}). For product wavefunctions, the probability density $P(a,\chi)$ is independent of $\rho$ (at least at the semiclassical level) and yields the expected classical behavior in the $\hbar \to 0$ limit.   

For more general $d$-dimensional minisuperspace models, one could employ the Klein-Gordon norm, which is defined in terms of an integral over a $(d-1)$-dimensional surface of the conserved Klein-Gordon current \cite{DeWitt}. Interpreting the coordinate minisuperspace field in the direction perpendicular to this surface as ``time,'' the Klein-Gordon norm and the associated probabilities are conserved. However, the Klein-Gordon norm is not positive definite and vanishes for real wavefunctions -- see \eg \cite{HP, Vilenkin5, Halliwell2} for discussions. In some simpler models, the problem of negative probabilities can be avoided by suitably choosing the $(d-1)$-dimensional integration surface (as shown by DeWitt in Ref.~\cite{DeWitt}), but it is not clear if this approach can be extended in the more general cases. 

Rather, we could seek to define a positive definite inner product based on the modulus of the wavefunction squared, $|\Psi|^2$, and an appropriate positive measure $\mu(a,\phi^i)$. For a certain operator ordering choice, the Hamiltonian becomes proportional to the laplacian $\nabla^2$ on the minisuperspace manifold, with respect to an appropriate metric of Lorentzian signature. Thus, the natural choice of measure $\mu(a, \phi^i)$ that ensures the Hermiticity of the Hamiltonian in this case is the volume element \cite{HP}. Imposing the Hermiticity of the Hamiltonian for generic operator orderings, we expect the measure to be modified to $\mu(a,\phi^i,\rho)$,\footnote{In fact, this measure could be obtained by generalizing the derivation given in Appendix A.5 of Ref.~\cite{Rpsi}.} where the function $\rho(a,\phi^i)$ parametrizes the ordering ambiguities in the higher $d$ cases at the semiclassical level.\footnote{Depending on the model, more than one function maybe needed to parametrize the possible ordering ambiguities.} It would be interesting to investigate whether  the resulting probability densities are independent of $\rho$, as in the one-dimensional cases. In addition, we would like to see if the expected classical probability distribution will be recovered in the $\hbar \to 0$ limit.  To this extend, it would be interesting to compare with the seminal work \cite{Halliwell2}, where probability distributions are constructed based on the decoherent histories approach to quantum mechanics.

Notice that as in some one-dimensional cases, it may turn out that the wavefunction is non-normalizable with respect to this inner-product in phenomenologically interesting higher $d$ cosmological models. So at best, we can use the wavefunction in these cases to define relative probabilities. 
For instance, it would be interesting to extend our approach to inflationary models such as the Starobinsky model, involving the scale factor and a scalar field with a  potential suitable to realize inflation. Finding the measure on the Hilbert space would allow us to define relative probabilities between configurations where the inflaton slowly rolls along the plateau of the potential and a configuration where the inflaton sits at the minimum of the potential. 
In principle, the structure of probability density would allow us to conclude whether initial conditions suitable for inflation are favored.

\section{Conclusion}
\label{cl}

Field redefinitions of the scale factor provide us with different prescriptions for implementing the no-boundary proposal and computing the wavefunction as a path integral. These wavefunctions satisfy different WDW equations that also admit ordering ambiguities occurring when quantizing the  Hamiltonians. At least at the semiclassical level, the inner products on the corresponding Hilbert spaces turn out to be equivalent, leading to universal predictions \cite{Rpsi}.

In this work we have shown that the quantum probability density reproduces the classical probability distribution in the  $\hbar \to 0$ limit, for a homogeneous and isotropic closed universe filled with a perfect fluid with equation of state $p_m=w\rho_m$, $w\in [-1,\, 1]$. When the index $w$ is in the range $(-1/3,1]$, the  cosmology has a finite lifetime, between a big bang and a big crunch, and the classical probability distribution can be normalized to unity. Likewise the corresponding quantum wavefunction is normalizable.   The quantum probability density is vanishingly small at the big bang and big crunch singularities, indicating that a singular universe could be  excluded at the quantum level. The quantum expectation value of the extrinsic curvature is finite provided that $w > 1/3$.

On the other hand, based on semiclassical considerations, we show that the quantum wavefunction is not normalizable when the prefect fluid index satisfies $-1\le w \le -1/3$, corresponding to accelerating cosmologies (for $w<-1/3$) with infinite duration. One however may use the wavefunction to define relative probabilities, with the classical behavior reproduced in the $\hbar \to 0$ limit.

\section*{Acknowledgements} 

The authors would like to thank Balthazar de  Vaulchier for discussions. A.K and N.T. would like to thank the Ecole Polytechnique for hospitality. The work of  A.K. is partially supported by the NTUA basic research grant no 65228100.

\end{document}